\documentclass[fleqn,usenatbib]{mnras}

\usepackage{newtxtext,newtxmath}

\usepackage[T1]{fontenc}

\DeclareRobustCommand{\VAN}[3]{#2}
\let\VANthebibliography\thebibliography
\def\thebibliography{\DeclareRobustCommand{\VAN}[3]{##3}\VANthebibliography}

\usepackage{graphicx}	
\usepackage{amsmath}	
\usepackage{amssymb}	

\usepackage{multicol}
\usepackage{color}
\usepackage{multibib}
\defcitealias{2017ApJS..230...15M}{MS17}
\defcitealias{2020MNRAS.tmp.1299S}{S20}
\defcitealias{2017PASA...34...58E}{E17}

\title[Stellar binary fraction indicators]{Binary Fraction Indicators in Resolved Stellar Populations and Supernova Type Ratios}

\author[E. R. Stanway et al.]{
E. R. Stanway,$^{1}$\thanks{E-mail: e.r.stanway@warwick.ac.uk}
J. J. Eldridge,$^{2}$
and A. A. Chrimes$^{1}$
\\
$^{1}$Department of Physics, University of Warwick, Gibbet Hill Road, Coventry, CV4 7AL, UK\\
$^{2}$Department of Physics, University of Auckland, Private Bag 92019, Auckland, New Zealand
}

\date{Accepted 2020 July 13. Received 2020 July 13; in original form 2020 June 15}

\pubyear{2020}

\begin{document}
\label{firstpage}
\pagerange{\pageref{firstpage}--\pageref{lastpage}}
\maketitle

\begin{abstract}
The binary fraction of a  stellar population can have pronounced effects on its properties, and in particular the number counts  of different massive star types, and the relative subtype rates of the supernovae which end their lives. Here we use binary population synthesis models with a binary fraction that varies with initial mass
to test the effects on resolved stellar pops and supernovae, and ask whether these can constrain the poorly-known binary fraction in different mass and metallicity regimes. We show that Wolf-Rayet star subtype ratios are valuable binary diagnostics, but require large samples to distinguish by models. Uncertainties in which  stellar models would be spectroscopically classified as Wolf-Rayet stars are explored. The ratio of thermonuclear, stripped envelope and other core-collapse supernovae may prove a more accessible test and upcoming surveys will be sufficient to constrain both the high mass and low mass binary fraction in the $z<1$ galaxy population.
\end{abstract}

\begin{keywords}
galaxies: stellar content -- binaries: general -- stars: evolution -- methods: numerical
\end{keywords}



\section{Introduction}

Stellar population synthesis models provide a framework through which observational data of stellar clusters, galaxies and galaxy populations can be interpreted \citep{1976ApJ...203...52T}. Identifying the properties of the observed population relies on matching the data to predictions determined by the age, mass, metallicity and other properties of the best-fitting model. Those predictions are sensitive to the assumed evolution of individual stars included in the synthesis model, which in turn depends on assumptions including the fraction of stars affected by binary evolution pathways.

While the majority of stellar population and spectral synthesis models currently in use neglect the role of stellar multiplicity \citep[e.g.][]{2003MNRAS.344.1000B,2005MNRAS.362..799M,2004A&A...425..881L}, there is an increasing recognition that its effects are important, particularly when interpreting young and distant stellar populations, or in determining the rates of transient objects \citep[e.g.][]{1991A&A...249..411V,1992ApJ...386..197T,1998A&A...333..557D,1998NewA....3..443V,2013MNRAS.433.1039Z,2014MNRAS.444.3466S,2016MNRAS.462.3302E,2019MNRAS.482..870E,2016MNRAS.456..485S,2016MNRAS.458L...6W,2016MNRAS.459.3614M,2016ApJ...826..159S,2018ApJ...869..123S,2020MNRAS.491.3479C,2020A&A...634A.134G,2020arXiv200207230Z}. The fraction of massive stars affected by a binary companion during their evolution is clearly substantial, and cannot be entirely neglected \citep{2012Sci...337..444S,2013A&A...550A.107S}. Nonetheless, implementing binary evolution pathways is both technically challenging and involves introducing additional assumptions for the binary fraction, and the distribution of initial binary parameters in the population, as well as the initial mass function (IMF). Constraints on these parameters have improved significantly in recent years \citep{2017ApJS..230...15M,2019ApJ...875...61M,2020arXiv200500014T}, but remain poor at low metallicities and outside the local Universe. 

In \citet{2019A&A...621A.105S} we began a programme to explore the impact of these uncertainties on stellar population predictions, by varying the initial mass function parameters assumed by the Binary Population and Spectral Synthesis \citep[BPASS, ][hereafter E17]{2017PASA...34...58E} model framework, while keeping the binary parameters fixed. In \citet[][hereafter S20]{2020MNRAS.tmp.1299S} we instead explored the impact of stellar binary population parameter uncertainties on the integrated light of stellar populations for a fixed IMF. In that work we considered both observational uncertainties on the binary parameters in the current v2.2 of BPASS, which are based on the analysis of \citet[][ hereafter MS17]{2017ApJS..230...15M}, and an extended grid of models in which the binary fraction as a function of mass is varied by an arbitrary amount. 

In parallel, recent work by \citet{2018ApJ...867..125D,2020arXiv200413040D} has explored the effect of both binary fraction and rotation on predictions for resolved stellar populations, using a custom set of models in which stars of all masses are assumed to share a common binary fraction. They identified the ratio of certain massive stellar types, and in particular the ratio of stripped-envelope, strong-wind, helium-atmosphere Wolf-Rayet (WR) stars to red supergiant (RSG) stars, as being sensitive to the binary fraction (and indeed rotational mixing) assumed.

Here we explore the impact of a mass-dependent binary fraction on both stellar type ratios and supernova type ratios using a grid of models with a wide range of possible initial mass-dependent binary fractions and metallicities. We explore whether binary fractions might be recovered from observations of resolved stellar populations in the local Universe, or of bright transients at cosmological distances. We also explore the impact on these interpretations of recent proposals that the minimum luminosity of WR stars identified spectroscopically may show a strong metallicity dependence.

The structure of this paper is as follows: In section \ref{sec:method} we introduce the model grid used here and discuss the alternate definitions of WR stars. In section \ref{sec:metal} we present the predictions of our models for continuously star forming populations as a function of metallicity. In section \ref{sec:redshift} we consider the binary fraction influence on supernova rates and the ratio between supernova types, assuming appropriate redshift histories for both star formation and its metallicity distribution. We evaluate the impact of WR definition and of binary fraction on these predictions, and consider whether upcoming projects will enable binary fraction to be evaluated observationally in future, in section \ref{sec:discussion}. Finally, we summarise our main conclusions in section \ref{sec:conc}.

\section{Method}\label{sec:method}

\subsection{Standard Models}\label{sec:bpass}

All models presented here are based on the Binary Population and Spectral Synthesis (BPASS) stellar population synthesis models \citep{2009MNRAS.400.1019E,2012MNRAS.419..479E,2016MNRAS.456..485S,2017PASA...34...58E}, specifically their v2.2.1 implementation \citep{2018MNRAS.479...75S}. This framework generates an evolving simple (i.e. coeval) stellar population in which the initial stellar masses are distributed according to a broken power law, and the binary fraction, initial period distribution and initial mass ratio distribution of stars are based on the distributions determined by \citetalias{2017ApJS..230...15M}. These were initially determined empirically for stars in five mass ranges and four initial period bins, and are interpolated onto the BPASS mass and period grid. Here we keep the initial mass function, initial period distribution and mass ratio distributions fixed in line with the BPASS v2.2 default, but vary the binary fraction with the logarithm of the mass of the primary star. 

\begin{figure}
	\includegraphics[width=\columnwidth]{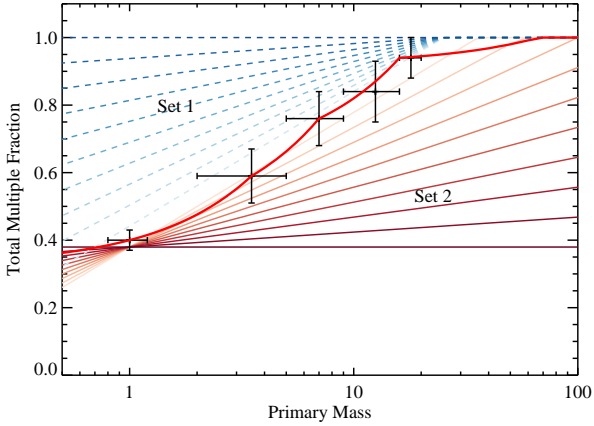}
    \caption{Multiple fractions tested in an experimental grid to examine possible observable signatures for binary populations. Each line indicates a model binary fraction distribution which either raises the binary fraction at low stellar mass (set 1, dashed lines) or lowers it at high mass (set 2, solid lines). Data points are drawn from \citetalias{2017ApJS..230...15M} and the thick red line indicates the fiducial model applied in BPASS v2.2.}
    \label{fig:fbase}
\end{figure}

As in \citetalias{2020MNRAS.tmp.1299S}, where the unresolved stellar populations derived from the same models are discussed, we define two sets of variant models. In set 1, the high mass binary star fraction (above 20\,M$_\odot$) is fixed at unity and the low mass binary fraction is permitted to vary from about 40 per cent at Solar mass up to unity. In set 2, the Solar mass binary star fraction is held fixed at about 40 per cent, but the high mass binary fraction is permitted to vary from its current estimate (near unity) down to 40 per cent. These sets of varying binary fractions are defined in Fig. \ref{fig:fbase} and discussed in detail in \citetalias{2020MNRAS.tmp.1299S}.

We note that this approach differs from and is complementary to that of \citet{2018ApJ...867..125D,2020arXiv200413040D} in which stars of all masses are deemed to share a common binary fraction, in conflict with the observed distributions in the local Universe. Since those papers addressed the relative numbers of massive stars, derived from a relatively narrow range of initial masses in young populations, their assumption of a constant binary fraction over that mass range is likely reasonable. However we expect the dependence on initial mass to affect any comparison with populations arising from lower mass stars - for example in the ratios of different supernova types as a function of metallicity or age, or their cosmic evolution (as discussed in section \ref{sec:redshift}).

The models presented here do not vary the distribution of initial binary separation and mass ratio due to computational constraints, but focus on the total binary fraction as a function of primary star mass. The effects of varying these parameters independently was explored for unresolved stellar populations by \citetalias{2020MNRAS.tmp.1299S}, and it is clear that the current observational constraints on separation and mass ratio permit a large range of possible models.  In the context of the work on resolved populations in this paper, the key question to be addressed is whether binary interactions alter the evolution of a system, thus changing its stellar type or supernova type at death. A system is more likely to interact if the stars begin their life in a close binary or if the mass ratio between primary and companion is near unity. Thus an increase in the total binary fraction has a similar effect to biasing the initial period distribution towards shorter periods, or to biasing the mass ratio towards twin systems. The default BPASS prescription for these is fixed based on observational constraints derived as a function of stellar mass by \citet{2017ApJS..230...15M}, and for massive stars already include a bias towards twin systems and short periods. Thus varying the overall binary fraction captures the majority of the behaviour for massive stars. For lower mass (e.g. Solar-type) stars, the distributions are broader and the observational constraints weaker, and so models in set 1 will be degenerate with models with larger mean separations or smaller mass ratios.

For each variant binary fraction versus mass distribution function, we calculate time-evolving stellar number counts for populations with an initial total stellar mass of $10^6$\,M$_\odot$ at 13 metallicities and 42 age steps, spaced logarithmically such that log(age/years)=$6.0+i\times\Delta$(age) ($i=0-41$) and the increment $\Delta$(age)=0.1. For each of these age steps, we assign each stellar model a type by luminosity, temperature and surface composition. 

Similarly we assign a type to each supernova identified based on the state of its progenitor at the end of its evolution. These classifications are described in \citet{2017PASA...34...58E}. Briefly, a star is considered to undergo a core-collapse supernova if it has undergone core carbon burning and has a CO-core mass $>1.38$\,M$_\odot$ at the end of its life. Its type is then determined by the chemical composition of the surface layers which will be ejected, and the remnant (if any) determined from the core mass after accounting for the supernova energy injection. The survival or disruption of the binary is determined probabilistically, given an assumed kick distribution. For stars with insufficient mass to undergo core collapse, the end state is deemed to be a white dwarf with the mass of the progenitor star's helium core at the end of its life.  Binary systems which survive to this point can show an increase in the white dwarf mass through mass transfer from a companion, or a merger of double white dwarfs through angular momentum loss due to gravitational wave radiation. Where either of these pathways result in a white dwarf with a total mass exceeding the Chandrasekhar limit, a thermonuclear, type Ia supernova is deemed to occur. The rates and delay time distributions of such explosive transients, as modelled in BPASS, are discussed in detail in \citet{2019MNRAS.482..870E} and are shown to be consistent with observational constraints.

\subsection{Wolf-Rayet definition}\label{sec:WRdef}

In the standard models described above, we have used the WR definitions laid out in \citetalias{2017PASA...34...58E} in which stars are identified as WR based primarily on their surface compositions. Stars are assumed to be identifiable as strong wind-driving, Wolf Rayet stars, rather than lower mass helium stars, if they have a luminosity exceeding log(L/L$_\odot$)$>4.9$. 

Recent work \citep{2020A&A...634A..79S} has argued on both observational and theoretical grounds that this simple constraint is insufficient. Instead, the luminosity constraint above which a star shows the spectral features classically identified as a  Wolf-Rayet may be metallicity dependent, scaling as $L^{WR}_\mathrm{spec}\propto Z^{-1}$. Stars below this threshold would show a blue, stripped star spectrum, but produce narrow line emission, rather than the strongly line-broadened emission associated with classical Wolf-Rayets.

To evaluate the impact of this proposal on the predicted number counts of stars by type, we recalculate the classification of stars in our models based on the relationship:
\[ \log_{10}\left(L^{WR}_\mathrm{spec}\right) = 4.9 - \log_{10}(Z/0.014).\] Only stars above this luminosity threshold are classified as WR.
These models are shown on figures with dotted lines, where appropriate. We do not expect this change to affect supernova rates, since these are determined by the structure and composition of the progenitor star, which is only weakly related to its 
stellar classification \citep[e.g.][]{2018PASA...35...49E}.

\section{Results}\label{sec:results}

\subsection{Trends with Metallicity}\label{sec:metal}

\subsubsection{Resolved stellar populations}

The metallicity of stars affects their wind strengths, radii, surface gravity and hence probability of undergoing binary interactions while on the main sequence or giant branch. Such interactions can lead to surface hydrogen stripping, rejuvenation and other processes which will change the classification of the stellar model. As a result, we expect (and observe) the ratio of different stellar types to depend on both binary fraction and metallicity. 

We calculate trends in stellar type number counts with metallicity for star forming stellar populations. In each case we assume that the composite stellar population (CSP) has been forming stars at a constant rate of 1\,M$_\odot$\,yr$^{-1}$ for 100 Myr, such that the number counts of most stellar types have stabilised, with the rate of stellar birth balanced by the rate of stellar death for massive stars. The long-lived low mass stellar population will continue to build up to much later ages, so we focus on the relatively massive stars which may be resolvable as individual stars beyond our immediate environs, and in particular on the Wolf-Rayet (WR) population of stripped-atmosphere stars.

\begin{figure}
	\includegraphics[width=\columnwidth]{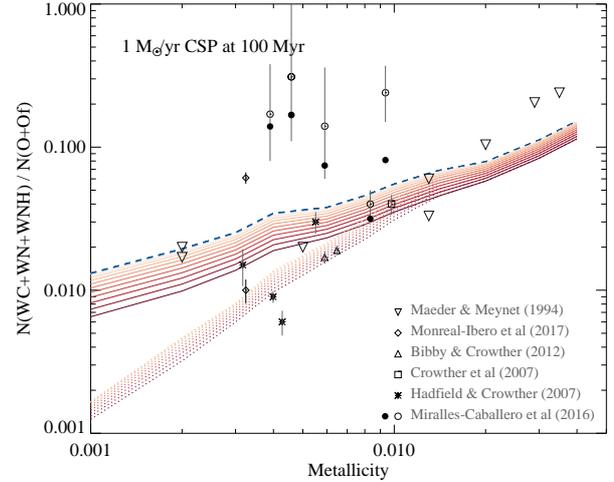}
    \caption{Wolf Rayet (WN + WC + WNH) to O star (O + Of, log(L/L$_\odot$)>4.9) ratio, as a function of metallicity and a range of binary fractions. Models are as colour coded in Fig \ref{fig:fbase}. Solid lines indicate a WR definition cut at log(L/L$_\odot$)>4.9, dotted lines are for a metallicity-dependent luminosity limit as discussed in section \ref{sec:WRdef}. Data points are from the references labelled \citep{1994A&A...287..803M,2017A&A...603A.130M,2012MNRAS.420.3091B,2007A&A...469L..31C,2007MNRAS.381..418H,2016A&A...592A.105M}. Filled symbols for \citet{2016A&A...592A.105M} indicate corrected values as discussed in Section \ref{sec:data_lims}.
}    \label{fig:wr_o}
\end{figure}

In Fig.\,\ref{fig:wr_o} we show the dependence of the WR to O-star ratio in such a population on metallicity and binary fraction. Unsurprisingly this ratio shows effectively no sensitivity to the binary fraction at low masses, with the models in set 1 indistinguishable at Solar metallicity. By contrast, the ratio is moderately dependent on the high mass binary fraction for our standard WR definition. Number count ratios yielded by the revised \citet{2020A&A...634A..79S} definition for WR stars show less dependence on binary fraction, but a stronger metallicity dependence than those using a uniform luminosity definition.

For context, we also show a compilation of observational data points reported for this ratio \citep{1994A&A...287..803M,2012MNRAS.420.3091B,2016A&A...592A.105M,2007MNRAS.381..418H,2007A&A...469L..31C}. In each case we use the values reported by the original authors without modification. Where authors give metallicity in the form of 12+log(O/H) we assume $Z=0.020$ corresponds to 12+log(O/H)=8.93 as appropriate for BPASS stellar evolution models \citep{2018MNRAS.477..904X,2017PASA...34...58E}. 
We note that this observational dataset is likely highly incomplete due to the  difficulty of resolving large samples of massive stars, determining their metallicity and classifying them reliably, and we discuss this further in Section \ref{sec:data_lims}. As a result of these uncertainties, the observational data show a large scatter and it is difficult to draw firm conclusions from the data. Nonetheless the models demonstrate that precision on the WR fraction significantly better than one per cent is needed to distinguish between binary fraction models at metallicities near Solar, where the ratio ranges from 0.078 at a massive star binary fraction of unity to 0.058 at a fraction of 40 per cent. 

\begin{figure}
	\includegraphics[width=\columnwidth]{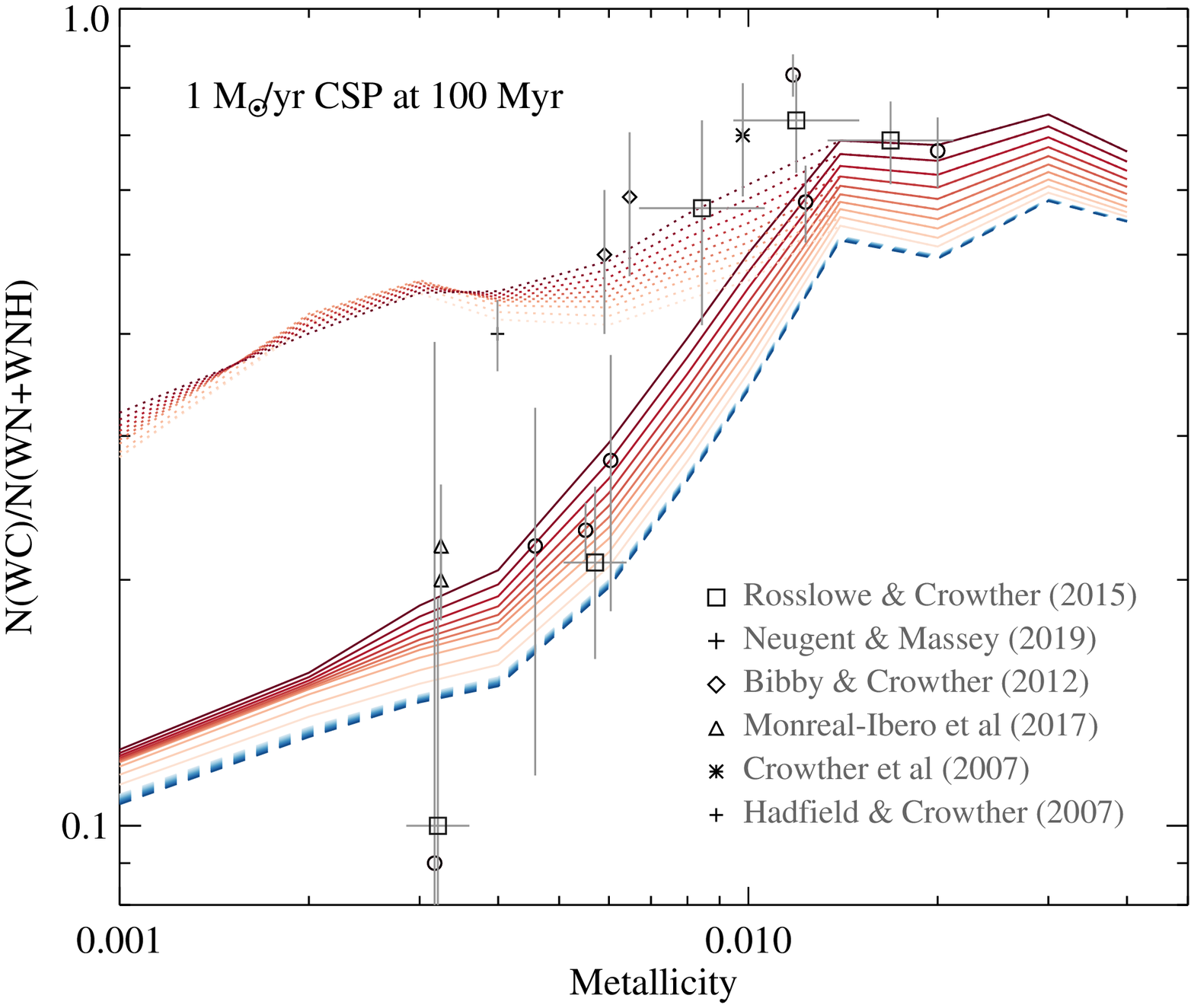}
    \caption{Wolf Rayet WN  to WC ratio as a function of metallicity for binary fractions as colour coded in Fig \ref{fig:fbase}. Dotted lines show the results for the revised WR definition. Data points are drawn from the literature \citep{2015MNRAS.447.2322R,2019Galax...7...74N,2012MNRAS.420.3091B,2017A&A...603A.130M,2007A&A...469L..31C,2007MNRAS.381..418H}.
    }
    \label{fig:wc_wn}
\end{figure}

A similar dependence on  metallicity in seen in the Wolf-Rayet subtype ratios shown in Fig \ref{fig:wc_wn}. The fraction of carbon-rich WC stars in the population (relative to nitrogen-rich WN stars and partially stripped WNH stars) declines sharply with either decreasing metallicity or increasing binary fraction when a uniform luminosity cut for WR stars is used. Introducing a metallicity dependence to the WR luminosity threshold has the effect of strongly reducing the dependence on both metallicity and binary fraction in this ratio. For comparison we show number counts for Galactic and Magellanic WR stars spanning a range of metallicities including the recent compillation from \citet{2015MNRAS.447.2322R}. 
While the uncertainties on these measurements are still very large 
they also appear to  disfavour the revised \citet{2020A&A...634A..79S} WR star definition. 

\begin{figure}
	\includegraphics[width=\columnwidth]{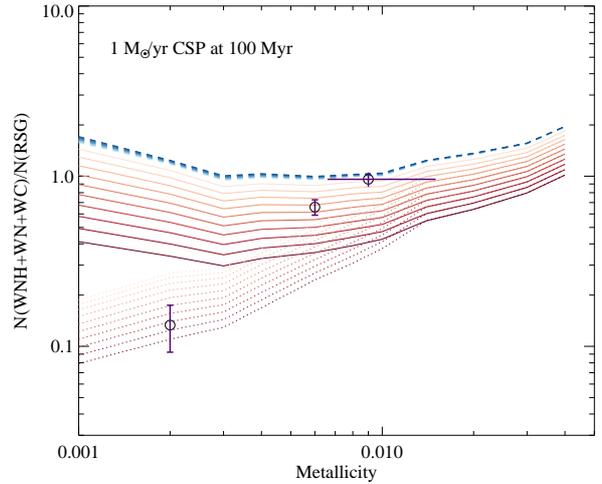}
    \caption{Wolf Rayet (WN + WC) to RSG (K+M, log(L/L$_\odot$)$>4.9$) ratio as a function of metallicity for binary fractions as colour coded in Fig \ref{fig:fbase}. Dotted lines show the ratio for the revised WR luminosity limit. Data points are drawn from the literature  \citep{2003AJ....126.2867M,2016AJ....152...62M,2018ApJ...863..181N}.}
    \label{fig:wr_rsg}
\end{figure}

Another observation that has been suggested as a sensitive probe of massive star populations  \citep[e.g.][]{1980A&A....90L..17M,2019Galax...7...74N,2016AJ....152...62M,2018ApJ...867..125D} is the WR to red supergiant (RSG, defined in our models as K or M type stars with log(L/L$_\odot$)$>$4.9) ratio. We show the metallicty dependence of this ratio in our models in figure \ref{fig:wr_rsg}. Interestingly, and unlike the previous two ratios considered, this quantity is only mildly dependent on metallicity when using our standard WR definition, but very strongly dependent on massive star binary fraction \citep[as also noted by][]{2018ApJ...867..125D}. This is a useful trait: the precise metallicity of stellar populations is often difficult to determine, particularly for more distant objects. Given the \citet{2020A&A...634A..79S} WR definition, the binary sensitivity remains but the ratio is now also metallicity dependent. Since the ratio is close to 1:1, small differences in the population ratio can be determined with relative ease - although the low number of objects in both classes still presents a problem. For comparison, we plot the ratio  for  M33  from \citet{2016AJ....152...62M} iand estimates for the SMC and LMC for which RSG data is drawn from \citet{2003AJ....126.2867M} and WR numbers from \citet{2018ApJ...863..181N}.  As demonstrated by \citet{2018ApJ...867..125D}, this line ratio is also dependent on the age of a simple stellar population, and so comparisons of Fig.~\ref{fig:wr_rsg} to data are not recommended for small starbursts or single-aged stellar clusters, but are likely to be robust in the larger populations such as galaxies which have been forming stars at a constant or slowly varying rate over $10^8$\,year timescales, such as M33. 

\begin{figure}
	\includegraphics[width=\columnwidth]{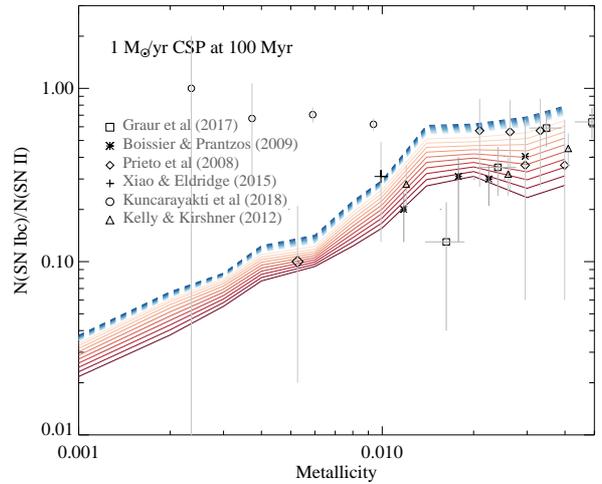}
    \caption{SN type II to SN type Ib/c ratio as a function of metallicity for binary fractions as colour coded in Fig \ref{fig:fbase}. Metallicity differences or uncertainties swamp binary fraction ones. For comparison, we show a compilation of data from the literature with representative uncertainties \citep{2008ApJ...673..999P,2009A&A...503..137B,2012ApJ...759..107K,2015MNRAS.452.2597X,2017ApJ...837..120G,2018A&A...613A..35K}.}
    \label{fig:supernovae}
\end{figure}

\subsubsection{Relative rates of supernovae}

While resolved stellar number counts such as those discussed above are promising binary fraction diagnostics, an alternative diagnostic can be derived from the manner in which these stars end their lives \citep[e.g.][]{2008MNRAS.384.1109E}. Stars which have been stripped or gained mass through binary interactions may produce explosions which are classified differently, shifting between hydrogen-rich (type II) and hydrogen-poor (type I) classes. Amongst these transients, the ratio of stripped-envelope to hydrogen-rich core-collapse supernovae shows promise as a diagnostic of binary fraction. As Fig.\,\ref{fig:supernovae} demonstrates, this ratio declines with decreasing metallicity, tracking the fraction of stripped envelope massive stars in the population. As before, we overplot these models with a representative sample of observational data, showing both the vast range of estimates in the literature, and the large uncertainties on current measurements.

\subsection{Cosmic Evolution}\label{sec:redshift}

The probes discussed above are sensitive to the massive star binary properties but relatively insensitive to the binary fraction amongst intermediate mass and Solar-type stars. To probe these, we need to identify sources or transients with low mass progenitors, and take into account the longer evolutionary lifetime of these stars. Hence we need to account for both a star formation and metallicity history over gigayear timescales. This is challenging for any one galaxy, but plausible on a volume-averaged scale where extensive work has gone in to determining both the star formation rate (SFR) density evolution \citep{2014ARA&A..52..415M} and the global metallicity evolution \citep{2006ApJ...638L..63L}\footnote{While other metallicity distribution estimates exist in the literature, the metallicity distribution of high redshift star formation remains very uncertain, and we retain this prescription for comparison with earlier work. As \citet{2020MNRAS.493L...6T} explored, this prescription allows the correct recovery of local transient rates.}. In this context we consider the cosmic evolution of supernova rates, considering both core collapse events (with massive progenitors) and thermonuclear detonations (with lower mass progenitors).

\begin{figure}
	\includegraphics[width=\columnwidth]{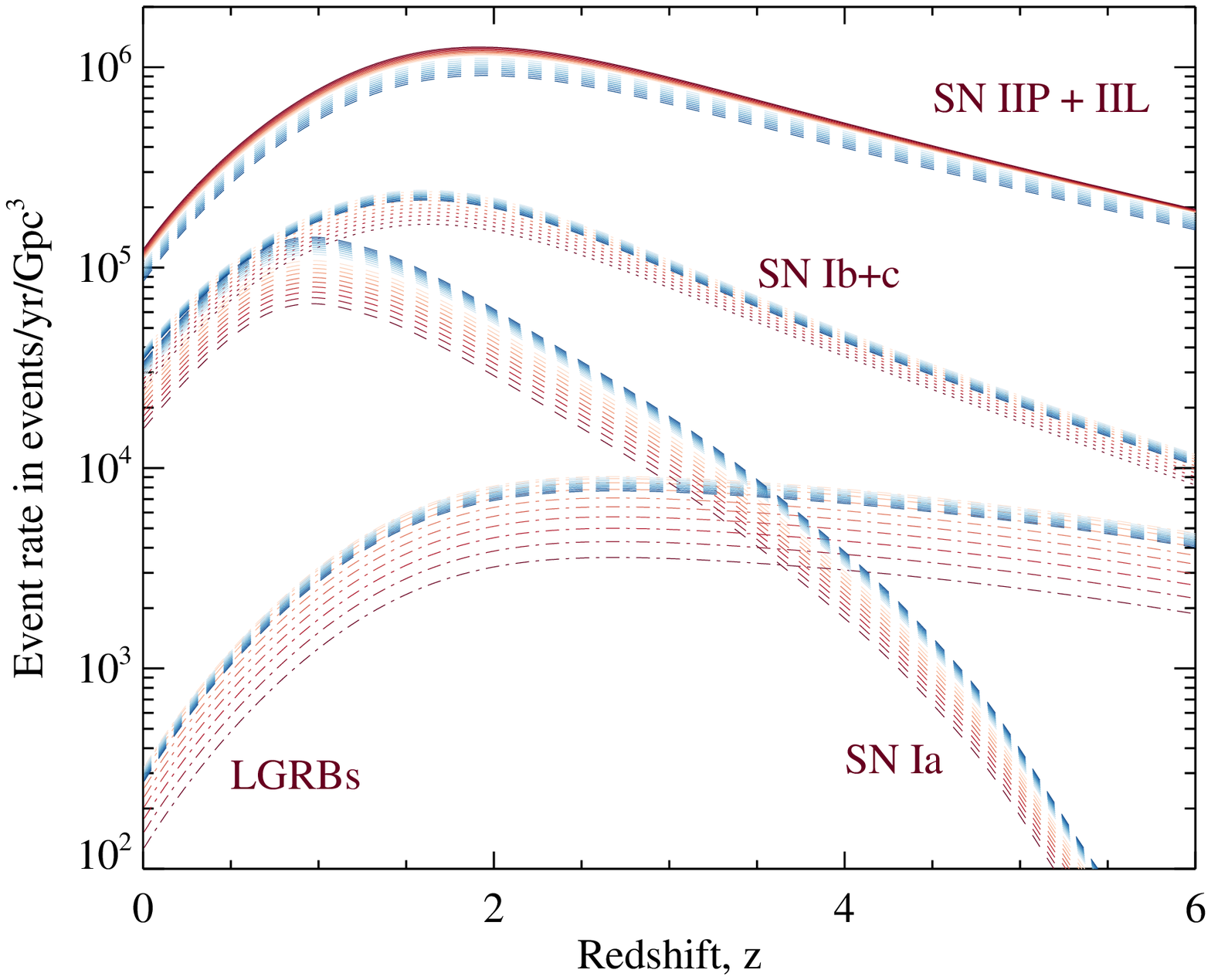}
	\includegraphics[width=\columnwidth]{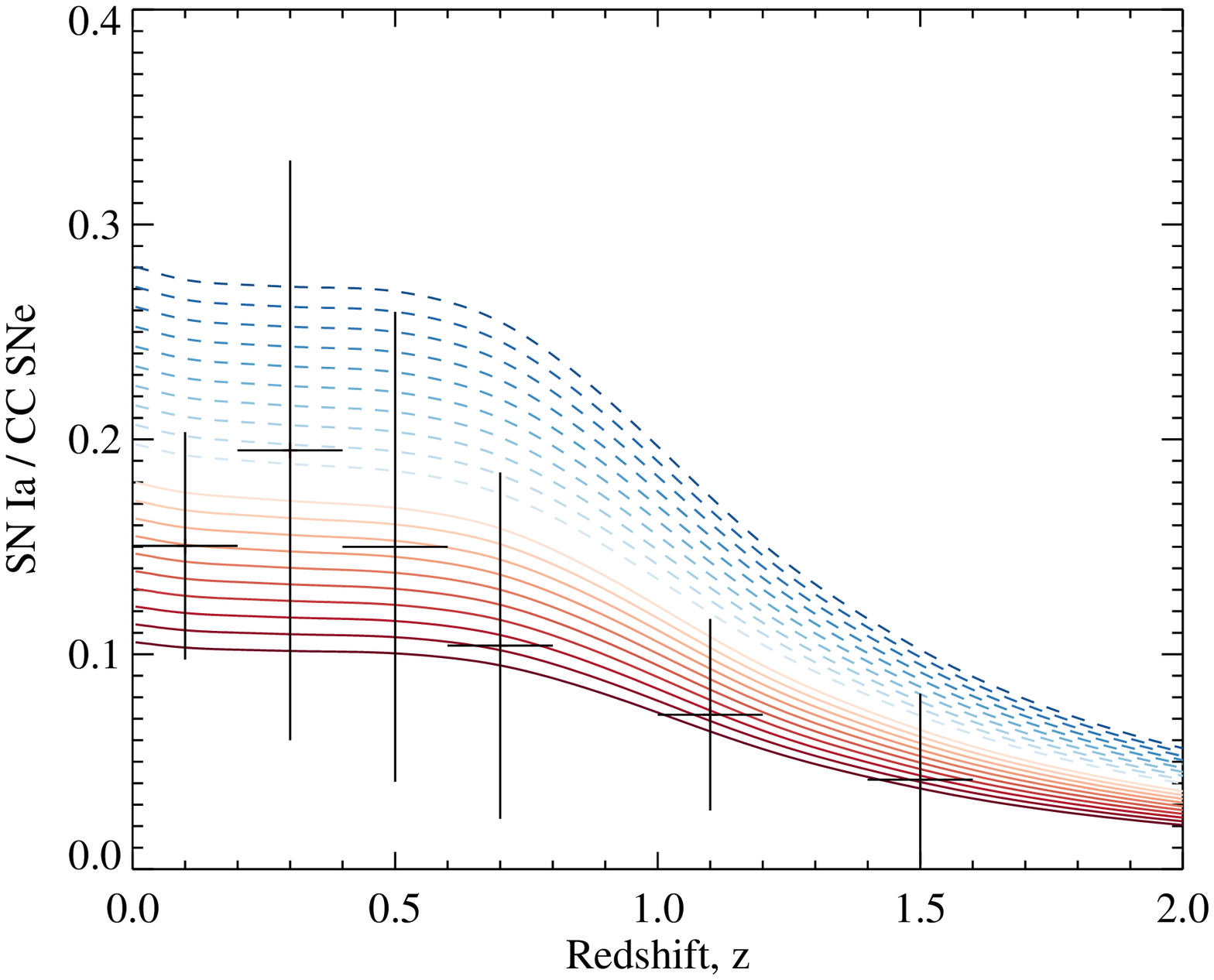}
	\includegraphics[width=\columnwidth]{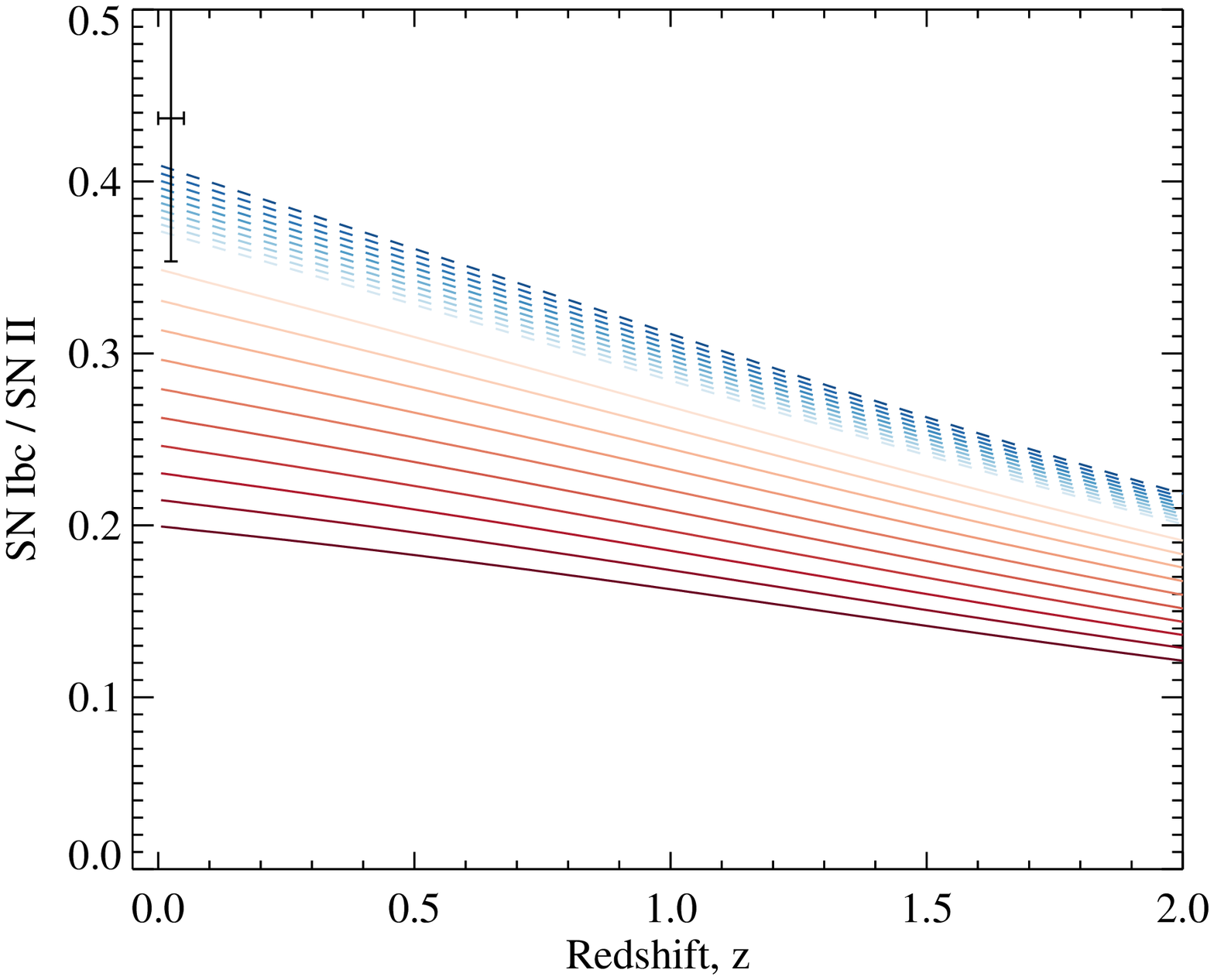}
    \caption{Cosmic Evolution in  volume-averaged supernova rates by type and type ratios, as a function of binary fraction model, assuming the redshift evolution prescriptions for SFR and Z adopted in \citet{2019MNRAS.482..870E}. Overplotted points show the current state of observational constraints, compiled as described in section \ref{sec:redshift}.}
    \label{fig:cosmic}
\end{figure}

We adopt the same cosmic evolution prescription for SFR and Z as \citet{2019MNRAS.482..870E} to calculate the star formation rate density distributed between different metallicities as a function of redshift. Using delay time distributions and event rates from our models, we calculate the resultant cosmic evolution of supernova rate per unit volume for each variant binary fraction distribution\footnote{We assume $\Omega_M=0.286$, $\Omega_\Lambda=0.714$, $h=0.696$.}. The results are shown in Fig. \ref{fig:cosmic}. The upper panel gives the evolution in the mean volumetric rate of each supernova type between $z=0$ and $z=6$. In the lower panels, the evolution in the ratio of different types is shown out to $z=2$ and compared to a compilation of observational data as described below. We note that the lines indicating Long Gamma Ray Bursts (LGRBs) include only the chemically homogeneous evolution pathway which dominates at the lowest metallicities, and neglects
pathways which operate at higher metallicity \citep[these may be included in later BPASS releases, see ][]{2020MNRAS.491.3479C}.

To constrain the observed ratio of thermonuclear type Ia rates to core collapse supernova rates, SN\,Ia \citep{2014PASJ...66...49O,2015A&A...584A..62C,2014AJ....148...13R,2012A&A...545A..96M,2014ApJ...783...28G,2012AJ....144...59P} and CCSN \citep[][and data compiled therein]{2014ApJ...792..135T,2012A&A...545A..96M,2009A&A...499..653B,2016A&A...594A..54P} volumetric rate data have been sorted into $\Delta z=0.2$ bins, and where one or more rate estimates for both types exist in the same redshift bin, their ratio is taken. 
For the stripped envelope supernova fraction we show the local rate ratio estimated from the LOSS survey \citep{2017PASP..129e4201S} for galaxies at $z<0.05$.

At low redshifts, a binary fraction close to unity is preferred for resolved studies of high mass stars, with some indication that a high binary fraction is also preferred for Solar-type stars at very low metallicity \citep{2017ApJS..230...15M,2019ApJ...875...61M}. In each case, however, the observational uncertainties on current survey data are too large to distinguish between binary fractions with any degree of reliability, or to evaluate the redshift evolution of these rates.

As Fig.\,\ref{fig:cosmic} demonstrates, the stripped envelope fraction amongst core collapse supernovae evolves linearly with redshift, reflecting the slow evolution in the metallicity of the underlying stellar population.  By contrast, the fraction of thermonuclear type Ia SNe relative to core collapse events remains near constant out to $z\sim0.7$ before declining sharply. This results primarily from the much longer delay times distribution of the type Ia events. These require the evolution of relatively low mass stars into white dwarfs, which then grow through binary interactions until the Chandrasekhar mass limit is reached.

\section{Discussion}\label{sec:discussion}

As we have demonstrated, both the types of massive stars and their eventual supernovae are sensitive to the presence of binary evolution pathways in the population. So are we approaching the point where resolved studies of massive stars may directly constrain the binary fraction of their underlying populations?

\subsection{Observations of stellar type ratios} \label{sec:data_lims}

The models presented in section \ref{sec:metal} are broadly consistent with the compilation of observational data shown, in terms of order of magnitude in number count ratios and underlying trends with metallicity. However Figures \ref{fig:wr_o}-\ref{fig:wr_rsg} also demonstrate that there are large variations in observational estimates of stellar type ratios. They also clearly indicate the very small number of measurements for which  estimates of metallicity and massive star number count ratios are available. Nonetheless, in certain ratios, and in particular the WR/RSG ratio, the uncertainties quoted on the data are already sufficiently small to interpret as binary fraction measurements Given these factors, it is important to assess the robustness and appropriateness of the samples against which we are comparing.

In their recent comprehensive survey of resolved massive stars in M31 and M33  \citet{2016AJ....152...62M} estimated that they were almost complete for Wolf-Rayet stars but were incomplete for RSGs in M31 and had identified only a few percent of the O stars present in the galaxies. In many of the observational samples reported, the completeness is still lower. 

The O star population is difficult to quantify  due to confusion in star forming regions and the typical brightness of individual stars. As a result, the number of O stars is often inferred from the ionizing photon flux inferred in a population, while the number and type of the Wolf-Rayet stars is inferred from fitting of mass-scaled templates to diagnostic spectral features\footnote{This approach is taken by all the data shown on Fig. \ref{fig:wr_o}, with individual WR stars only resolved in the very closest objects such as in parts of the LMC and SMC, and O star numbers always derived indirectly.}. As a result, dusty stars may be undercounted, as may the hottest stars which radiate primarily in the ultraviolet. It is also an inconvenient fact that known Wolf-Rayet stars have luminosities that scatter over two orders of magnitude \citep{2006A&A...449..711C} and so determining whether any individual ionized region has been irradiated by one star or many is challenging \citep[see e.g.][]{2015MNRAS.447.2322R}. This leads to a large scatter in the WR/O star number ratios reported, ranging from those which rely on clear identification of individual stars (incomplete) to those based entirely on inference from unresolved populations (heavily model and metallicity dependent). To illustrate the scale of these effects, in Fig \ref{fig:wr_o} we show two sets of points for the data of \citet{2016A&A...592A.105M}: open circles indicate the values given by the original authors as inferred from fitting unresolved stellar populations, filled circles indicate values using the original WR numbers but modifying the inferred O star count to account for the generally lower ionizing flux to O star number ratio in the BPASS models. As the figure demonstrates, this increases the number of O stars inferred and brings this estimate closer into line with other estimates at similar metallicity. Nonetheless ratios inferred from this data set remain high compared to other data.

Each data set presents its own challenges to interpret. In several cases, no uncertainty is given on the published number ratios, and where possible this is inferred to give error bars on Fig. \ref{fig:wr_o} from Poisson number count uncertainties on the inferred population. These have decreased with publication date as the number of detected sources per galaxy has risen. However Poisson uncertainties do not account for systematic uncertainties in the underlying models used to infer the numbers, which can easily be of order a few tenths of a dex and thus span the model parameter space here. A fully consistent comparison between models and data would require the model completeness and calibration calculations to be undertaken using BPASS or a comparable code which incorporates binary evolution pathways.

Where Wolf-Rayet stars are identified, either individually or through spectral fitting, they are typically classified into WC or WN types by the strength of carbon features in the spectrum. Thus many of the uncertainties which affect the data in Fig. \ref{fig:wr_o} also affect Fig. \ref{fig:wc_wn}, with the added challenge that subdividing the small Wolf-Rayet population adds to the Poisson uncertainties. Again, it is not always clear whether systematic modelling uncertainties are incorporated in the reported error bars for these data, and is likely that the true uncertainty on most of the data encompasses the full span of the models. In this context, it is interesting to note that above a metallicity of about half Solar, the data appear to favour models with low fractions of massive binaries, which are inconsistant with those observed in the local Universe \citep{2012Sci...337..444S,2014ApJS..215...15S,2017ApJS..230...15M}. This may indicate that the number of WN stars in local galaxies is being underestimated using current template fitting techniques.

In comparison to the ratios discussed above, data for the WR/RSG ratio shown in Fig. \ref{fig:wr_rsg} is very sparse in the literature: while RSG and WR populations have been studied separately in local group galaxies, it is rarely possible to evaluate whether the same regions have been surveyed in each case, the metallicity of the region being considered and the relative levels of completeness in the samples. In the figure, we have shown estimates for the SMC and LMC for which RSG data is drawn from \citet{2003AJ....126.2867M} and WR numbers from \citet{2018ApJ...863..181N}. While these works originate from the same team, they are derived from very different imaging surveys, with different spatial coverage. As a result the ratio can be compromised by the inclusion or omission of bright star forming regions, or particularly young regions in one survey which may be omitted from the other, or conversely by a more extended, more mature stellar population. The third data point on Fig. \ref{fig:wr_rsg} is that for M33  in which \citet{2016AJ....152...62M} identified and spectroscopically confirmed 211 resolved WR stars and 220 RSGs and estimated that the survey was near complete for WR stars, and may also be complete for RSGs. This data point (at $\sim$0.5 Solar metallicity) is entirely consistent with the high binary fractions inferred for massive stars elsewhere in the local Universe. Unfortunately, the metallicity of this system is rather uncertain, with the 1\,$\sigma$ error range admitting models with binary fractions of about 70 per cent or higher at 30\,M$_\odot$. This point resulted from a substantial, multi-year campaign, but demonstrates the potential for constraints on the stellar binary fraction from large nearby galaxies.

In short, where data based on counting of individual stars is available (primarily in the SMC, LMC and perhaps M33), the data may be used with caution. Where number counts are inferred from unresolved populations, stellar population model dependence and completeness must be carefully considered.

\subsection{Constraints from star number counts}

The extant observational data cannot distinguish between WR definitions in either the WR/O or the WR/RSG ratio, but hints that the revised luminosity limit suggested by \citet{2020A&A...634A..79S} cannot reproduce the trend in WC/WN ratio with luminosity, for which our original log(L/L$_\odot$)=4.9 luminosity limit, independent of metallicity, provides a good match. If there is indeed a strong metallicity dependence in the luminosity limit for Wolf-Rayet spectroscopic identification, then the apparent discrepancy between the data and these predictions would suggest that the mass-loss rates, and especially their scaling with metallicity, in the BPASS stellar evolution models need to be revised. This question will be revisited in future work, since there is growing evidence that the mass-loss rates for WR stars and RSGs may need to be revised generally \citep[e.g.][]{2015PASA...32...15Y,2017MNRAS.470.3970Y,2020MNRAS.492.5994B,2020ApJ...889...44N}.

Setting aside the definition question, and focussing on our standard fixed-luminosity selection, the WR-to-O star ratio ranges from almost 8 per cent at a massive star binary fraction of unity to 6 per cent at a fraction of 40 per cent. As a result, distinguishing these populations at any reasonable degree of confidence would require an observed Wolf-Rayet population well over ten thousand objects - far more than the total number of currently known WR stars in the Milky Way and its satellites. Thus it is unlikely that this ratio will be determined to sufficient precision in any given galaxy to act as a strong constraint on the binary population. 

Since binary processes are, at least in part, responsible for stripping the envelopes of stars which might otherwise evolve into WR stars, the WR/RSG ratio shows promise for evaluating the binary fraction in local galaxies in the near future. As \citet{2016AJ....152...62M} demonstrated, this ratio can be determined in large nearby galaxies with a high degree of precision, given sufficient observational time and effort. The ratio is relatively insensitive to metallicity, mitigating an often-substantial degeneracy in the fitting of any data, and shows a strong sensitivity to the binary fraction in massive stars. 

\subsection{Future prospects for star count observations}

Given the model-dependence of indirectly-inferred number counts, there is a clear preference for sensitive observations of resolved stars that allow counting of sources down to a luminosity limit of at least log(L/L$_\odot$)=4.9. In this context, it is worth considering what observations future instrumentation may enable in this area.

Science cases for the upcoming class of Extremely Large Telescopes (ELTs) include the detailed study of resolved stellar populations beyond the local group. The MICADO instrument on the European ELT\footnote{EELT, https://www.eso.org/sci/facilities/eelt/}, for example, would expect to resolve and detect stars down to the horizontal branch at the distance of the Centaurus group ($\sim$4.6\,Mpc) in five hours of integration, and so should produce complete catalogues for red supergiants \citep{2012PASP..124..653G}. The  fields of view expected for ELT instruments are expected to be less than a square arcminute (in some cases, significantly less) and while this is suitable for mapping distant galaxies, will require large mosaics to map Local Group objects.

However, like many of the planned ELT instruments, MICADO is optimised to operate in the near-infrared, where adaptive optics can be most effectively deployed. As a result, it is unlikely to provide any information on Wolf-Rayet and other luminous blue supergiant stars, for which near-ultraviolet imaging is preferred.  Optical spectroscopy provides an alternate method for identifying Wolf-Rayet stars, as described as above, but the first-light spectrograph on the ELT is not expected to be sufficiently blue-sensitive.

In the nearer term, resolved stellar populations may also be accessible to the {\it James Webb Space Telescope} ({\it JWST}) and an early release science programme in this area has been approved in Cycle 0 \citep{2017jwst.prop.1334W}. As is the case for the ELTs, JWST is a near-infrared optimised observatory with a small field of view. It will reach comparable sensitivities to the ELTs due to lying above the atmosphere, but suffers from a larger point spread function. As a result, confusion is likely to be an issue for observations at significant distances, while large mosaics will be necessary to map nearby galaxies. An optimal application for JWST may be study of individual star forming regions or complexes, for which the metallicity, age and binary fraction can be determined simultaneously, in contrast to the constant star formation case considered here. 

The effort to identify and map Wolf-Rayet stars, however, is unlikely to benefit significantly from either JWST or the ELTs due to their near-infrared optimisation. For these, the current and ongoing effort to identify these sources from integral field spectroscopy and optical photometry is unlikely to be improved upon before the construction of a blue-sensitive, large aperture observatory such as the proposed LUVOIR \footnote{https://asd.gsfc.nasa.gov/luvoir/}. Continuing this work, with a goal of highly complete spectroscopic follow-up, wherever possible of individually resolved sources, is essential if constraints on the binary fraction are to be obtained from stellar type number count ratios.

It should also be noted that while these instruments are not optimised for mapping the large angular scales subtended by Local Group galaxies, analysis of the resolved stellar populations in more compact and distant objects may allow  average ratios may be derived for larger samples of galaxies as a function of metallicity which will shed light on these populations. As with any observation, it will be crucial to map different stellar populations, fit any spectra and determine metallicities self-consistently and for stars drawn from the same spatial regions, before comparison can be made to model predictions such as those presented here.

\subsection{Constraints from supernova observations}

All the number count ratios involving WR stars are, however, relatively insensitive to the binary fraction in low mass binary stars in the population, as might be expected. The strongest diagnostic of low mass binaries studied here is the ratio of SN\,Ia to core-collapse supernovae. As Fig.~\ref{fig:cosmic} demonstrates, distinguishing between high mass star binary fractions requires precision on the SN\,Ia or SN Ibc fraction of about 1 per cent at $z=0$ and becomes progressively more difficult at higher redshifts. A similar precision is needed to constrain binary fraction as a function of metallicity, as seen in \ref{fig:supernovae} in which the data uncertainties are dominated by corrections for completeness in calibration or follow up. Since stripped envelope supernovae are often harder to classify from lightcurves than hydrogen-rich SN II, many of the estimates shown are likely to be lower limits.

While demanding, the required precision promises to be eminently achievable with the upcoming Legacy Survey of Space and Time (LSST) at the Vera Rubin Observatory. LSST will carry out a deep, high cadence survey of the transient sky, expecting to find of order $10^5$ type Ia supernovae per year, and a comparable number of core collapse events \citep[][see chapter 11]{2009arXiv0912.0201L}. The majority of these will lie in the range $z=0.2-1$, an interval over which the ratio of event type is expected to change significantly - as Fig.~\ref{fig:cosmic} shows. Given the expected rate of events, if all could be accurately typed, measurements would be possible of the supernova type ratios in ten redshift bins at about 1 per cent precision - sufficient to distinguish between high and low binary fractions at both ends of the mass function. 
With lower numbers, of only about 1000 SNe per $\Delta z=0.1$ bin, the number of measured SNe\,Ia is expected to be 200, SNe\,Ibc about 240 and SN\,II about 560 giving 7 per cent uncertainty, 6 per cent uncertainty and 4 per cent uncertainty respectively on measured rates from simple Poisson statistic arguments - these then need to be corrected for observational biases. With 10,000 SNe per bin, the Poisson uncertainties drop to 2, 2 and 1 per cent, sufficient to identify the binary fraction to within $\pm1$ model on our current grid. This will be true for CCSN out to $z=0.5$ in 1 year \citep{2009JCAP...01..047L}. Higher redshifts may be accessible through wider redshift bins, while extended data as the survey continues will enable narrower bins to be used, probing more details such as the metallicity history of the galaxy evolution. 

We note that this assumes redshift uncertainties are smaller than the bin size.  At this redshift range, this should be possible in the majority of host galaxies through photometric redshift determination. It also assumes that supernovae can be accurately typed by their lightcurves in the absence of large-scale spectroscopy \citep[expected to be true, ][]{2009arXiv0912.0201L}.

We have also assumed that the same binary fraction applies at all metallicities, and that the same distribution of period and mass ratio applies at all binary fractions.  These are more difficult to quantify or justify as assumptions and further studies with a more extensive suite of models will be required to evaluate the extent to which the joint posterior probability distribution of these parameters can be determined. Intriguingly, the wide area and deep limits of the LSST data will enable lensed supernovae to be observed at much higher redshifts. \citet{2020MNRAS.491.2447R} estimated that up to 120 lensed supernovae at $z\sim5-7$ could be detected by the LSST Wide Deep Fast survey, with more sources at intermediate redshift. While the precision in any type ratio derived from this higher redshift population would necessarily be large, it will provide an important test of the metallicity distribution assumed for high redshift star formation in this model.

In very local examples, identified in LSST or other survey data on well studied local galaxies, it might be possible to determine both the supernova type ratio and WR/RSG ratio, at least for large galaxies. A simultaneous analysis of the SN type ratios and WR/RSG ratios for the same sample of galaxies would be a powerful diagnostic tool. 
This combination yields a diagnostic grid in binary fraction vs metallicity for $Z>0.002$. Again a precision of about 1 per cent is required to distinguish between models in SN type ratio, while a lower precision (about 10 per cent) is sufficient in the harder-to-measure stellar type ratio, and this is still likely to be challenging for the current and next generation of facilities.

\section{Conclusions}\label{sec:conc}

Analysis of the type statistics of massive stars has the potential to constrain the fraction of binary stars in stellar populations. However the degree of precision required is significantly higher than that obtained by current surveys.

Adopting the metallicity dependence suggested by \citet{2020A&A...634A..79S} for the minimum luminosity of classical Wolf-Rayet stars significantly changes both the metallicity and binary fraction dependence of Wolf-Rayet number type ratios. Both the WR/O and WR/RSG ratios become more strongly metallicity dependent, while the WC/WN ratio becomes less so, in mild conflict with recent observational evidence. More  data on these line ratios (drawn from large, complete sample of resolved stars, or potentially from the integrated light of well-aged stellar clusters) are needed before the new WR definition is adopted. We note that \citeauthor{2020A&A...634A..79S} do not argue that stripped helium stars at luminosities between log(L/L$_\odot$)=4.9 and their metallicity dependent limit do not exist or do not affect their surroundings, but rather than they would not show the characteristic spectral features indicative of strong stellar winds.

The synergy between the capabilities of upcoming telescopes in the fields of resolved stellar populations (e.g. JWST, ELTs) and supernova rates (e.g. LSST) has the capacity to constrain the binary fraction as a function of metallicity and even redshift. LSST's vast dataset will likely allow both the high and low mass binary fractions to be determined to a high degree of precision, with some constraints on its metallicity evolution if the cosmic evolution of supernova type ratios can be measured with sufficient precision. This relies on reliable typing of supernovae, either photometrically or spectroscopically. 

We have focussed here on the effect of varying the total binary fraction at a given mass. Since stars in wide binaries (log(initial period/days)$>$4) are unlikely to interact in a Hubble time, and are treated as single stars in BPASS, this variation is degenerate with fixing the binary fraction, but instead biasing its period distribution towards closer binaries. Distinguishing between these scenarios is likely to be far harder, in the absence of spectroscopic period determinations for large numbers of distant stellar populations - beyond the capabilities of even planned telescopes. Constraining the period and mass ratio distributions based on very local stars is likely to remain necessary for some time to come.

\section*{Acknowledgements}

ERS  recieved support from United Kingdom Science and Technology Facilities Council (STFC) grant number ST/P000495/1 and ST/T000406/1. AAC was supported by STFC studentship 1763016. JJE acknowledges support from the University of Auckland and the Royal Society Te Ap\={a}rangi of New Zealand under the Marsden Fund.
BPASS would not be possible without the computational resources of the University of Auckland's NZ eScience Infrastructure (NeSI) Pan Cluster and the University of Warwick's Scientific Computing Research Technology Platform (SCRTP).   

\section*{Data Availability}

The model data reported here is tabulated in the appendix and will be made available via the BPASS websites - bpass.auckland.ac.uk or warwick.ac.uk/bpass. 






%
\appendix
\section{Binary Fraction Distributions}

The binary fraction, $f_\mathrm{bin}$, in these models is defined as a function of mass by a parameterisation:
\[
f_\mathrm{bin}(M_i) = \mathrm{min}([1.0, A\times\log_{10}(M_i)+f_\mathrm{bin}(1\,M_\odot)]),
\]
where $M_i$ is the initial mass of the primary or single star in Solar masses and $A$ is a constant selected to produce the two model sets shown in Fig.~\ref{fig:fbase}. Values of $A$, and $f_\mathrm{bin}(1\,M_\odot)$ used here are given in Table \ref{tab:a1}, together with resultant values for $f_\mathrm{bin}(30\,M_\odot)$.

We also provide numerical values for the set 2 (massive star binary fraction) models in Figs.~\ref{fig:wr_o}-\ref{fig:supernovae} in Tables \ref{tab:a2}, \ref{tab:a3} and \ref{tab:a4}.

\begin{table}
    \caption{Parameters of the binary fraction functions used in this work.}
    \label{tab:a1}
    \centering
    \begin{tabular}{ccccc}
       Set & Model & A &  $f_\mathrm{bin}(1\,M_\odot)$  & $f_\mathrm{bin}(30\,M_\odot)$\\
        \hline\hline
     1 & 0 &  0.399  & 0.442 &  1.000       \\
       & 1 &  0.354  & 0.504 &  1.000       \\
       & 2 &  0.310  & 0.566 &  1.000       \\
       & 3 &  0.266  & 0.628 &  1.000       \\
       & 4 &  0.222  & 0.690 &  1.000       \\
       & 5 &  0.177  & 0.752 &  1.000       \\
       & 6 &  0.133  & 0.814 &  1.000       \\
       & 7 &  0.089  & 0.876 &  1.000       \\
       & 8 &  0.044  & 0.938 &  1.000       \\
       & 9 &  0.000  & 1.000 &  1.000       \\
    \hline       
     2 & 0  &      0.399  &  0.380 &  0.969 \\
       & 1  &      0.354  &  0.380 &  0.903 \\
       & 2  &      0.310  &  0.380 &  0.838 \\
       & 3  &      0.266  &  0.380 &  0.772 \\
       & 4  &      0.222  &  0.380 &  0.707\\
       & 5  &      0.177  &  0.380 &  0.641 \\
       & 6  &      0.133  &  0.380 &  0.576 \\
       & 7  &      0.089  &  0.380 &  0.510 \\
       & 8  &      0.044  &  0.380 &  0.445\\
       & 9  &      0.000  &  0.380 &  0.380\\
       \hline
    \end{tabular}
\end{table}

\begin{table*}
    \caption{Predicted number count ratios for model set 2, assuming 100 Myr of constant star formation at 1\,M$_\odot$\,yr$^{-1}$. Metallicities shown are $Z=0.001, 0.002, 0.003, 0.004$. }
    \label{tab:a2}
    \centering
    \begin{tabular}{c|cccccccc}
       & \multicolumn{3}{c}{$\log(L^{WR}/L_\odot)>4.9$)} & & & \multicolumn{3}{c}{$\log(L^{WR}/L_\odot)>4.9-\log(Z/0.014)$}\\
 $Z$=0.001  Model  &  N(WR/O) &     N(WC/WN) &    N(WR/RSG) &    N(SNIbc/SNII)   & &   N(WR/O) &     N(WC/WN) &    N(WR/RSG) \\
        0 &   0.01266 &   0.11228 &   1.43995 &   0.03382  &  &  0.00164 &   0.28243 &   0.18695 \\
        1 &   0.01224 &   0.11546 &   1.29543 &   0.03280  &  &  0.00166 &   0.28277 &   0.17594 \\
        2 &   0.01169 &   0.11815 &   1.15310 &   0.03155  &  &  0.00166 &   0.28657 &   0.16394 \\
        3 &   0.01105 &   0.11958 &   1.01833 &   0.03023  &  &  0.00163 &   0.29347 &   0.14988 \\
        4 &   0.01037 &   0.12002 &   0.89345 &   0.02882  &  &  0.00157 &   0.29675 &   0.13533 \\
        5 &   0.00966 &   0.12050 &   0.77981 &   0.02740  &  &  0.00151 &   0.30030 &   0.12208 \\
        6 &   0.00892 &   0.12109 &   0.67602 &   0.02598  &  &  0.00145 &   0.30432 &   0.10998 \\
        7 &   0.00815 &   0.12182 &   0.58083 &   0.02456  &  &  0.00139 &   0.30889 &   0.09887 \\
        8 &   0.00736 &   0.12274 &   0.49322 &   0.02315  &  &  0.00132 &   0.31416 &   0.08866 \\
        9 &   0.00652 &   0.12393 &   0.41233 &   0.02173  &  &  0.00125 &   0.32028 &   0.07922  \\
        \\
 $Z$=0.002         
      Model   &      N(WR/O) &     N(WC/WN) &    N(WR/RSG) &    N(SNIbc/SNII)   & &   N(WR/O) &     N(WC/WN) &    N(WR/RSG) \\
        0   &   0.01864 &   0.13597 &   1.08192 &   0.06036  &  &  0.00463 &   0.40966 &   0.26868 \\
        1   &   0.01799 &   0.14053 &   0.98022 &   0.05866  &  &  0.00465 &   0.41291 &   0.25310 \\
        2   &   0.01719 &   0.14480 &   0.87988 &   0.05653  &  &  0.00458 &   0.42117 &   0.23461 \\
        3   &   0.01628 &   0.14693 &   0.78440 &   0.05407  &  &  0.00445 &   0.42333 &   0.21418 \\
        4   &   0.01531 &   0.14763 &   0.69455 &   0.05131  &  &  0.00426 &   0.42072 &   0.19321 \\
        5   &   0.01429 &   0.14843 &   0.61190 &   0.04855  &  &  0.00406 &   0.41772 &   0.17390 \\
        6   &   0.01325 &   0.14940 &   0.53565 &   0.04582  &  &  0.00386 &   0.41430 &   0.15609 \\
        7   &   0.01216 &   0.15058 &   0.46509 &   0.04310  &  &  0.00365 &   0.41038 &   0.13961 \\
        8   &   0.01104 &   0.15204 &   0.39961 &   0.04040  &  &  0.00343 &   0.40584 &   0.12432 \\
        9   &   0.00987 &   0.15393 &   0.33867 &   0.03771  &  &  0.00321 &   0.40052 &   0.11008 \\
        \\
 $Z$=0.003           
    Model   &    N(WR/O) &     N(WC/WN) &    N(WR/RSG) &    N(SNIbc/SNII)   & &   N(WR/O) &     N(WC/WN) &    N(WR/RSG) \\
        0   &   0.02438 &   0.14922 &   0.86775 &   0.07949  &  &  0.00811 &   0.44829 &   0.28864 \\
        1   &   0.02355 &   0.15502 &   0.79020 &   0.07714  &  &  0.00809 &   0.45552 &   0.27141 \\
        2   &   0.02259 &   0.16060 &   0.71524 &   0.07483   & &  0.00799 &   0.46316 &   0.25302 \\
        3   &   0.02151 &   0.16417 &   0.64310 &   0.07235  & &  0.00779 &   0.46516 &   0.23301 \\
        4   &   0.02034 &   0.16640 &   0.57426 &   0.06947  &  &  0.00752 &   0.46320 &   0.21242 \\
        5   &   0.01912 &   0.16898 &   0.51053 &   0.06660  &  &  0.00724 &   0.46098 &   0.19334 \\
        6   &   0.01785 &   0.17205 &   0.45137 &   0.06374  &  &  0.00695 &   0.45849 &   0.17564 \\
        7   &   0.01654 &   0.17575 &   0.39632 &   0.06090  &  &  0.00664 &   0.45568 &   0.15916 \\
        8   &   0.01518 &   0.18030 &   0.34496 &   0.05807  &  &  0.00633 &   0.45250 &   0.14378 \\
        9   &   0.01376 &   0.18602 &   0.29693 &   0.05526  &  &  0.00600 &   0.44886 &   0.12941 \\
        \\
 $Z$=0.004           
    Model   &    N(WR/O) &     N(WC/WN) &    N(WR/RSG) &    N(SNIbc/SNII)   & &   N(WR/O) &     N(WC/WN) &    N(WR/RSG) \\
        0   &   0.03303 &   0.15739 &   0.90329 &   0.10990  &  &  0.01375 &   0.41644 &   0.37612 \\
        1   &   0.03186 &   0.16385 &   0.82659 &   0.10621  &  &  0.01359 &   0.42544 &   0.35266 \\
        2   &   0.03055 &   0.16982 &   0.75235 &   0.10268  &  &  0.01333 &   0.43343 &   0.32817 \\
        3   &   0.02910 &   0.17422 &   0.68042 &   0.09918  &  &  0.01293 &   0.43804 &   0.30227 \\
        4   &   0.02754 &   0.17744 &   0.61143 &   0.09544  &  &  0.01244 &   0.43963 &   0.27607 \\
        5   &   0.02593 &   0.18119 &   0.54711 &   0.09174  &  &  0.01192 &   0.44137 &   0.25163 \\
        6   &   0.02426 &   0.18561 &   0.48701 &   0.08809  &  &  0.01140 &   0.44335 &   0.22880 \\
        7   &   0.02253 &   0.19091 &   0.43074 &   0.08448  &  &  0.01085 &   0.44560 &   0.20742 \\
        8   &   0.02075 &   0.19739 &   0.37793 &   0.08091  &  &  0.01029 &   0.44819 &   0.18736 \\
        9   &   0.01891 &   0.20548 &   0.32828 &   0.07738  &  &  0.00970 &   0.45119 &   0.16849 \\
   \end{tabular}
\end{table*}

 \begin{table*}
    \caption{Predicted number count ratios for model set 2, assuming 100 Myr of constant star formation at 1\,M$_\odot$\,yr$^{-1}$. Metallicities shown are $Z=0.006, 0.008, 0.010, 0.014$. }
    \label{tab:a3}
    \centering
    \begin{tabular}{ccccccccc}
       & \multicolumn{3}{c}{$\log(L^{WR}/L_\odot)>4.9$)} & & & \multicolumn{3}{c}{$\log(L^{WR}/L_\odot)>4.9-\log(Z/0.014)$}\\
$Z$=0.006           
    Model   &    N(WR/O) &     N(WC/WN) &    N(WR/RSG) &    N(SNIbc/SNII)   & &    N(WR/O) &     N(WC/WN) &    N(WR/RSG) \\
        0   &   0.03669 &   0.20763 &   0.88075 &   0.12237  &  &  0.02168 &   0.41032 &   0.52048 \\
        1   &   0.03557 &   0.21617 &   0.81000 &   0.11805  &  &  0.02135 &   0.42072 &   0.48619 \\
        2   &   0.03431 &   0.22466 &   0.74195 &   0.11457  &  &  0.02091 &   0.43080 &   0.45206 \\
        3   &   0.03292 &   0.23188 &   0.67612 &   0.11144  &  &  0.02033 &   0.43852 &   0.41748 \\
        4   &   0.03144 &   0.23865 &   0.61362 &   0.10834   & &  0.01967 &   0.44492 &   0.38395 \\
        5   &   0.02991 &   0.24643 &   0.55526 &   0.10531  &  &  0.01899 &   0.45205 &   0.35262 \\
        6   &   0.02832 &   0.25547 &   0.50063 &   0.10234  &  &  0.01829 &   0.46006 &   0.32330 \\
        7   &   0.02668 &   0.26612 &   0.44939 &   0.09943  &  &  0.01756 &   0.46912 &   0.29581 \\
        8   &   0.02499 &   0.27885 &   0.40124 &   0.09657  &  &  0.01681 &   0.47945 &   0.26996 \\
        9   &   0.02323 &   0.29431 &   0.35590 &   0.09377  &  &  0.01603 &   0.49135 &   0.24563 \\
        \\
$Z$=0.008            
    Model   &    N(WR/O) &     N(WC/WN) &    N(WR/RSG) &    N(SNIbc/SNII)   & &    N(WR/O) &     N(WC/WN) &    N(WR/RSG) \\
        0   &   0.04444 &   0.27950 &   0.92304 &   0.19284   & &  0.03173 &   0.44081 &   0.65904 \\
        1   &   0.04315 &   0.29024 &   0.85337 &   0.18438   & &  0.03112 &   0.45337 &   0.61538 \\
        2   &   0.04173 &   0.30130 &   0.78627 &   0.17626   & &  0.03039 &   0.46619 &   0.57256 \\
        3   &   0.04013 &   0.31117 &   0.72033 &   0.16831   & &  0.02948 &   0.47728 &   0.52912 \\
        4   &   0.03839 &   0.32033 &   0.65668 &   0.16039   & &  0.02844 &   0.48702 &   0.48645 \\
        5   &   0.03659 &   0.33085 &   0.59693 &   0.15262   & &  0.02736 &   0.49800 &   0.44638 \\
        6   &   0.03473 &   0.34307 &   0.54074 &   0.14500   & &  0.02625 &   0.51048 &   0.40871 \\
        7   &   0.03280 &   0.35745 &   0.48781 &   0.13752   & &  0.02510 &   0.52480 &   0.37322 \\
        8   &   0.03082 &   0.37459 &   0.43786 &   0.13018    &&  0.02391 &   0.54139 &   0.33972 \\
        9   &   0.02876 &   0.39539 &   0.39065 &   0.12296   & &  0.02268 &   0.56084 &   0.30806 \\
        \\
 $Z$=0.010         
    Model   &    N(WR/O) &     N(WC/WN) &    N(WR/RSG) &    N(SNIbc/SNII)   & &    N(WR/O) &     N(WC/WN) &    N(WR/RSG) \\
        0   &   0.05348 &   0.35872 &   0.93747 &   0.26018   & &  0.04450 &   0.46473 &   0.78008 \\
        1   &   0.05186 &   0.37154 &   0.87110 &   0.24813   & &  0.04340 &   0.47869 &   0.72893 \\
        2   &   0.05013 &   0.38456 &   0.80718 &   0.23651   & &  0.04218 &   0.49273 &   0.67920 \\
        3   &   0.04820 &   0.39654 &   0.74393 &   0.22487   & &  0.04077 &   0.50542 &   0.62917 \\
        4   &   0.04617 &   0.40834 &   0.68310 &   0.21309   & &  0.03925 &   0.51756 &   0.58074 \\
        5   &   0.04408 &   0.42179 &   0.62578 &   0.20146   & &  0.03769 &   0.53125 &   0.53509 \\
        6   &   0.04194 &   0.43727 &   0.57167 &   0.18997   & &  0.03609 &   0.54679 &   0.49200 \\
        7   &   0.03975 &   0.45529 &   0.52051 &   0.17862   & &  0.03446 &   0.56460 &   0.45126 \\
        8   &   0.03750 &   0.47651 &   0.47207 &   0.16741   & &  0.03278 &   0.58520 &   0.41268 \\
        9   &   0.03519 &   0.50188 &   0.42613 &   0.15634   & &  0.03106 &   0.60930 &   0.37610 \\
        \\
$Z$=0.014           
    Model   &    N(WR/O) &     N(WC/WN) &    N(WR/RSG) &    N(SNIbc/SNII)   & &    N(WR/O) &     N(WC/WN) &    N(WR/RSG) \\
        0   &   0.06676 &   0.54251 &   1.14112 &   0.51089   & &  0.06676 &   0.54251 &   1.14112 \\
        1   &   0.06496 &   0.55593 &   1.06599 &   0.48315   & &  0.06496 &   0.55593 &   1.06599 \\
        2   &   0.06306 &   0.56858 &   0.99468 &   0.45631   & &  0.06306 &   0.56858 &   0.99468 \\
        3   &   0.06090 &   0.58021 &   0.92296 &   0.42848   & &  0.06090 &   0.58021 &   0.92296 \\
        4   &   0.05849 &   0.59279 &   0.85184 &   0.40073   & &  0.05849 &   0.59279 &   0.85184 \\
        5   &   0.05601 &   0.60703 &   0.78450 &   0.37379   & &  0.05601 &   0.60703 &   0.78450 \\
        6   &   0.05348 &   0.62330 &   0.72068 &   0.34763   & &  0.05348 &   0.62330 &   0.72068 \\
        7   &   0.05088 &   0.64205 &   0.66011 &   0.32222   & &  0.05088 &   0.64205 &   0.66011 \\
        8   &   0.04822 &   0.66390 &   0.60254 &   0.29752   & &  0.04822 &   0.66390 &   0.60254 \\
        9   &   0.04549 &   0.68968 &   0.54775 &   0.27350   & &  0.04549 &   0.68968 &   0.54775 \\
   \end{tabular}
\end{table*}

\begin{table*}
   \caption{Predicted number count ratios for model set 2, assuming 100 Myr of constant star formation at 1\,M$_\odot$\,yr$^{-1}$. Metallicities shown are $Z=0.020, 0.030, 0.040$. In this range, the new WR definition does not affect the number type ratios.}
    \label{tab:a4}
    \centering
    \begin{tabular}{ccccccccc}
        & \multicolumn{3}{c}{$\log(L^{WR}/L_\odot)>4.9$)} & & & \multicolumn{3}{c}{$\log(L^{WR}/L_\odot)>4.9-\log(Z/0.014)$}\\
 $Z$=0.020           
    Model   &    N(WR/O) &     N(WC/WN) &    N(WR/RSG) &    N(SNIbc/SNII)   & &    N(WR/O) &     N(WC/WN) &    N(WR/RSG) \\
        0   &   0.07766 &   0.51162 &   1.25668 &   0.52002   & &  0.07766 &   0.51162 &   1.25668 \\
        1   &   0.07611 &   0.52435 &   1.17784 &   0.49206   & &  0.07611 &   0.52435 &   1.17784 \\
        2   &   0.07445 &   0.53831 &   1.10341 &   0.46675   & &  0.07445 &   0.53831 &   1.10341 \\
        3   &   0.07243 &   0.55282 &   1.02803 &   0.44228   & &  0.07243 &   0.55282 &   1.02803 \\
        4   &   0.07019 &   0.56797 &   0.95421 &   0.41868   & &  0.07019 &   0.56797 &   0.95421 \\
        5   &   0.06788 &   0.58500 &   0.88430 &   0.39578   & &  0.06788 &   0.58500 &   0.88430 \\
        6   &   0.06550 &   0.60426 &   0.81802 &   0.37356   & &  0.06550 &   0.60426 &   0.81802 \\
        7   &   0.06304 &   0.62624 &   0.75509 &   0.35199    &&  0.06304 &   0.62624 &   0.75509 \\
        8   &   0.06049 &   0.65154 &   0.69526 &   0.33104    &&  0.06049 &   0.65154 &   0.69526 \\
        9   &   0.05787 &   0.68099 &   0.63832 &   0.31068   & &  0.05787 &   0.68099 &   0.63832 \\
        \\
 $Z$=0.030           
    Model   &    N(WR/O) &     N(WC/WN) &    N(WR/RSG) &    N(SNIbc/SNII)   & &    N(WR/O) &     N(WC/WN) &    N(WR/RSG) \\
        0   &   0.11031 &   0.59539 &   1.46330 &   0.55540   & &  0.11031 &   0.59539 &   1.46330 \\
        1   &   0.10820 &   0.60631 &   1.37930 &   0.51366   & &  0.10820 &   0.60631 &   1.37930 \\
        2   &   0.10598 &   0.61799 &   1.30148 &   0.47414   & &  0.10598 &   0.61799 &   1.30148 \\
        3   &   0.10322 &   0.63105 &   1.22140 &   0.43595   & &  0.10322 &   0.63105 &   1.22140 \\
        4   &   0.10013 &   0.64481 &   1.14176 &   0.39916   & &  0.10013 &   0.64481 &   1.14176 \\
        5   &   0.09697 &   0.66007 &   1.06603 &   0.36375    &&  0.09697 &   0.66007 &   1.06603 \\
        6   &   0.09374 &   0.67707 &   0.99395 &   0.32961    &&  0.09374 &   0.67707 &   0.99395 \\
        7   &   0.09043 &   0.69615 &   0.92527 &   0.29670   & &  0.09043 &   0.69615 &   0.92527 \\
        8   &   0.08704 &   0.71770 &   0.85974 &   0.26494   & &  0.08704 &   0.71770 &   0.85974 \\
        9   &   0.08356 &   0.74224 &   0.79716 &   0.23427   & &  0.08356 &   0.74224 &   0.79716 \\
        \\
 $Z$=0.040           
    Model   &    N(WR/O) &     N(WC/WN) &    N(WR/RSG) &    N(SNIbc/SNII)   & &  N(WR/O) &     N(WC/WN) &    N(WR/RSG) \\
        0   &   0.15025 &   0.55661 &   1.84268 &   0.65091   & &  0.15025 &   0.55661 &   1.84268 \\
        1   &   0.14754 &   0.56365 &   1.73927 &   0.60591   & &  0.14754 &   0.56365 &   1.73927 \\
        2   &   0.14457 &   0.57226 &   1.64298 &   0.56132   & &  0.14457 &   0.57226 &   1.64298 \\
        3   &   0.14084 &   0.58234 &   1.54363 &   0.51676   & &  0.14084 &   0.58234 &   1.54363 \\
        4   &   0.13666 &   0.59323 &   1.44449 &   0.47317   & &  0.13666 &   0.59323 &   1.44449 \\
        5   &   0.13238 &   0.60525 &   1.35024 &   0.43103   & &  0.13238 &   0.60525 &   1.35024 \\
        6   &   0.12801 &   0.61857 &   1.26054 &   0.39030   & &  0.12801 &   0.61857 &   1.26054 \\
        7   &   0.12355 &   0.63340 &   1.17507 &   0.35088   & &  0.12355 &   0.63340 &   1.17507 \\
        8   &   0.11899 &   0.65003 &   1.09354 &   0.31273   & &  0.11899 &   0.65003 &   1.09354 \\
        9   &   0.11433 &   0.66879 &   1.01569 &   0.27578   & &  0.11433 &   0.66879 &   1.01569 \\
    \end{tabular}
 \end{table*}


\bsp	
\label{lastpage}
\end{document}